\documentclass{article}
\usepackage{spconf}
\usepackage{cite}
\usepackage{stfloats}
\usepackage[cmex10]{amsmath}
\usepackage{amssymb}
\usepackage{graphicx}
\usepackage{multirow}
\usepackage{enumitem}
\usepackage{lipsum}
\usepackage{array}

\newcolumntype{C}[1]{>{\centering}m{#1}}
\usepackage[tight,footnotesize]{subfigure}
\usepackage{booktabs}
\usepackage{caption}
\date{}

\usepackage{enumitem}

\usepackage{url}

\usepackage{hyperref} 

\begin{document}
	
	\onecolumn
	
	\begin{description}[labelindent=0cm,leftmargin=3cm,rightmargin=3cm,style=multiline]
		
		\item[\textbf{Citation}]{M. Alfarraj, H. Di, and G, AlRegib, "Multiscale fusion for seismic geometric attribute enhancement." SEG Technical Program Expanded Abstracts 2017. Society of Exploration Geophysicists, 2017. 2310-2314.}

		\item[\textbf{DOI}]{\url{https://doi.org/10.1190/segam2017-17750698.1}}
		
		\item[\textbf{Review}]{Date of presentation: 26 September. 2017}
		
		\item[\textbf{Data and Codes}]{\href{https://github.com/olivesgatech/Multiscale-fusion-for-seismic-geometric-attribute-enhancement}{\underline{GitHub Link}}}

		\item[\textbf{Bib}] {@incollection{alfarraj2017multiscale,\\
  title={Multiscale fusion for seismic geometric attribute enhancement},\\
  author={Alfarraj, Motaz and Di, Haibin and AlRegib, Ghassan},\\
  booktitle={SEG Technical Program Expanded Abstracts 2017},\\
  pages={2310--2314},\\
  year={2017},\\
  publisher={Society of Exploration Geophysicists}\\
}}

		
		\item[\textbf{Contact}]{\href{mailto:motaz@gatech.edu}{motaz@gatech.edu}  OR \href{mailto:alregib@gatech.edu}{alregib@gatech.edu}\\ \url{http://ghassanalregib.com/} \\ }
	\end{description}
	
	\thispagestyle{empty}
	\newpage
	\clearpage
	\setcounter{page}{1}
	
	\twocolumn
	
\title{Multiscale fusion for seismic geometric attribute enhancement}
\name{Motaz Alfarraj, Haibin Di, and Ghassan AlRegib}
\address{Center for Energy and Geo Processing (CeGP) \\ School of Electrical and Computer Engineering \\ Georgia Institute of Technology}
\maketitle

\begin{abstract}
In this abstract, we propose a multiscale fusion technique to enhance seismic geometric attributes, such as dip and curvature, which are very sensitive to noise present in seismic data. For a give seismic section, first, we construct a Gaussian pyramid that allows us to generate the seismic attribute at different resolutions (scales). Then, all attributes at the different scales are fused together to form the proposed multiscale enhanced attribute. Applications to the 3D seismic dataset over the Great South Basin in New Zealand demonstrate that the proposed method is capable of improving both the resolution and noise robustness of the first-order dip and the second-order curvature attributes, compared to existing methods and algorithm. Such improvement indicates the great potential of our multiscale fusion technique for enhancing the quality of more multitrace seismic attributes, such as coherence, flexure, and GLCM.
\end{abstract}

\section{Introduction}
\begin{figure}[ht!]
		\centering
		\includegraphics[width=\linewidth]{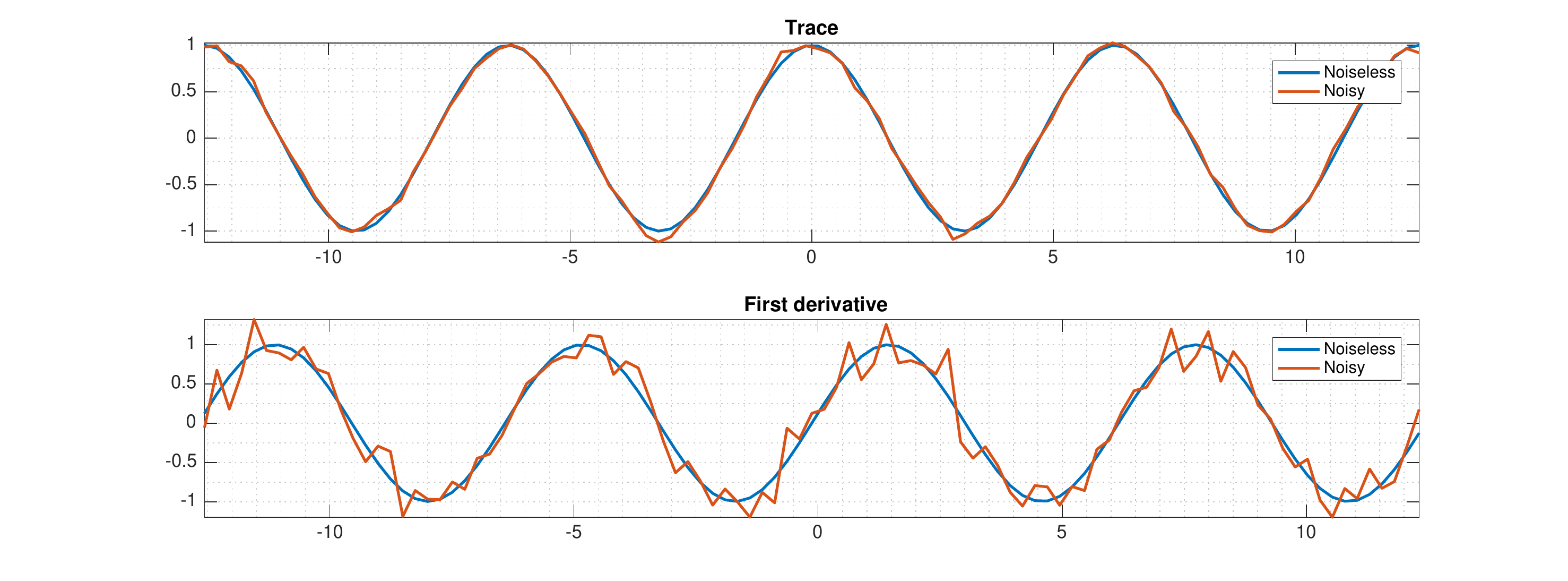}
		\caption{A clean trace, a noisy trace, and their derivatives.}
		\label{figure:derivative}
\end{figure}

Seismic geometric attribute (e.g., coherence, dip, and curvature) have been routinely employed in the process of 3D seismic interpretation for the purpose of structure delineation and fault detection. However, such attributes are known to be very sensitive to noise traditionally present in seismic data, especially the first-order dip and the second-order curvature attributes. For example, the dip attribute could be estimated in various approaches as summarized by \cite{marfurt2006robust}. One of these approaches is the phase dip (\cite{barnes1996theory}) which is considered most computational efficient based on the theory of the complex seismic trace analysis (\cite{taner1979complex}). However, such algorithm involves a suite of $1^\textrm{st}$ order differentiation of seismic signals, which in turn boosts the noise levels as illustrated in Figure \ref{figure:derivative}. Noisy attributes are traditionally enhanced using larger analysis windows or by post-processing filtering (\cite{barnes2007tutorial,di2014new}). In practice, while reducing the noise and the associated artifacts, these techniques also undesirably compromise the resolution of the processed attribute, rendering small-scale features invisible from the attribute images.

To overcome such limitation, in this work we propose implementing multiscale fusion to enhance the resolution of seismic geometric attributes (e.g., dip and curvature). The proposed enhancement workflow is threefold. First, a Gaussian pyramid is constructed to represent a seismic section at different scales, each of which exploits seismic features at different resolutions. Second, seismic attribute extraction is performed on each scale of the pyramid. Third, attributes from all scales of the pyramid are fused to form a multiscale attribute. In the result section, we test the proposed workflow on the 3D seismic dataset over the Great South Basin in New Zealand.

\section*{Methods}
In image processing, multiscale analysis methods refer to the representation of an image at different resolution levels, each of which comprises features of the same scale in terms of size. These methods have been widely used in image processing for a broad spectrum of applications such as image coding, compression, analysis and detection (e.g.  \cite{adelson1980image,burt1983laplacian}). Pyramid representation is one of the classical methods used in multiscale analysis which gave rise to scale-space theory and other multiscale signal representations. The Gaussian pyramid, the Laplacian pyramid and steerable pyramid (\cite{adelson1984pyramid,simoncelli1995steerable}), are examples of such pyramid representations.

In this work, we utilize the Gaussian pyramid as an efficient multiscale analysis tool to exploit features of different sizes in a seismic section. For a 2D seismic section $f_0[m,n]\in \mathbb{R}^{M\times N}$, the $K$-scale Gaussian pyramid is constructed as follows. $f_0[m,n]$ is set as scale 0 of the pyramid which represents the full resolution scale. Then, scale 1 of the pyramid, $f_1[m,n]$, is computed by blurring $f_0[m,n]$ with a Gaussian kernel then downsampling by a factor of 2. The remaining scale are generated in a similar fashion. The Gaussian pyramid construction is formulated in equation \ref{eqn:pyramid}. 
\begin{equation}
f_{i}[m,n] = \sum_{\forall k,l \in \mathbf{S}} g[k,l]f_{i-1}[2m+k,2n+l],
\label{eqn:pyramid}
\end{equation}
where $g[k,l]$ is a Gaussian kernel defined over a region $\mathbf{S}$ as follows, 
\begin{equation}
g[m,n] = \frac{1}{2\pi\sigma^2}\exp\left(-\frac{m^2+n^2}{2\sigma^2}\right),~~\forall [m,n] \in \mathbb{S}
\end{equation}

\begin{figure}[ht!]
		\centering
		\includegraphics[width=\linewidth]{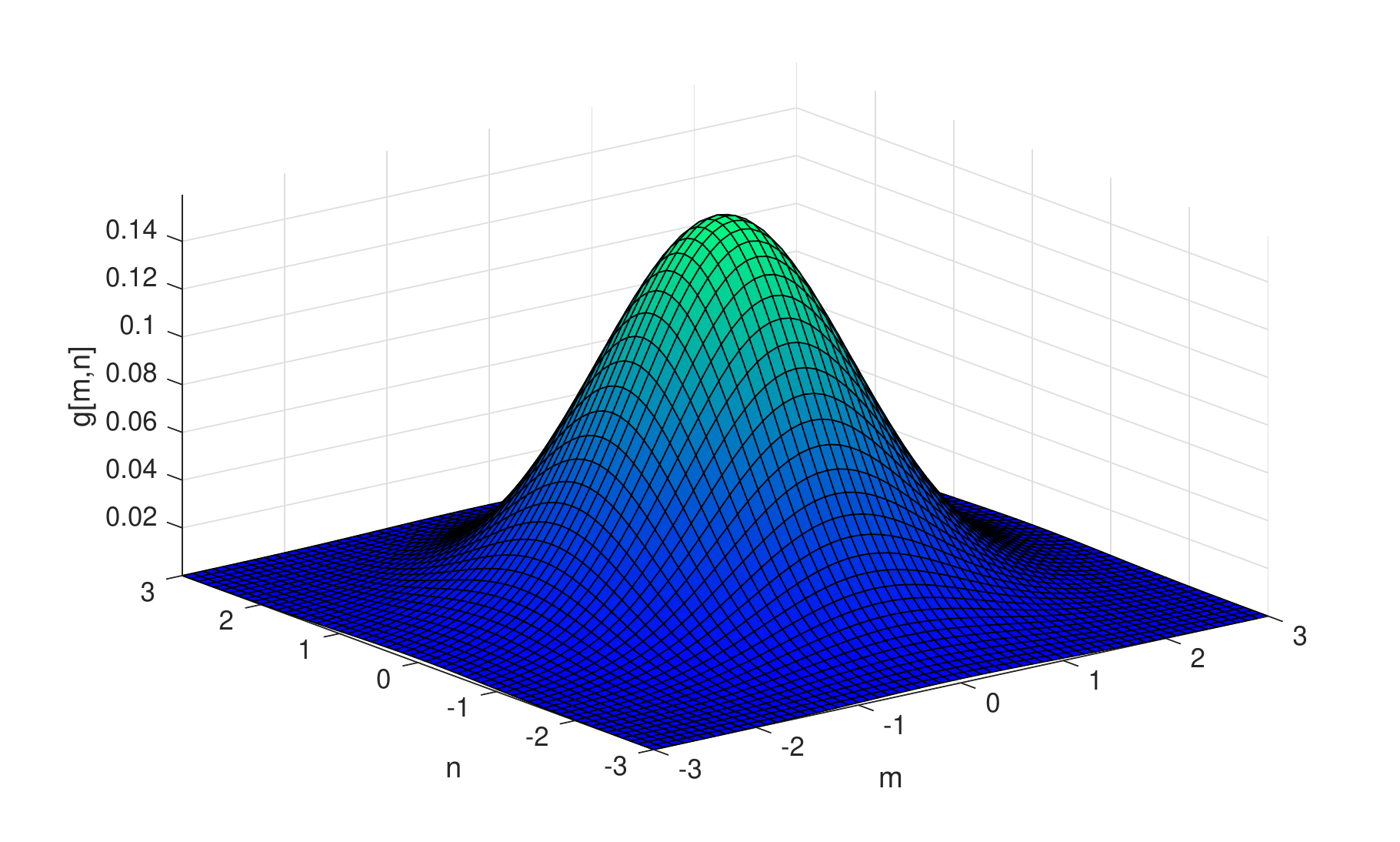}
		\caption{An example of a Gaussian kernel.}
		\label{figure:gaussain}
\end{figure}

A typical choice of $\mathbf{S}$ is a rectangular support of size $5\times 5$. An example of a Gaussian Kernel with unit standard deviation is shown in Figure \ref{figure:gaussain}. 

\begin{figure}[ht!]
		\centering
		\includegraphics[width=\linewidth]{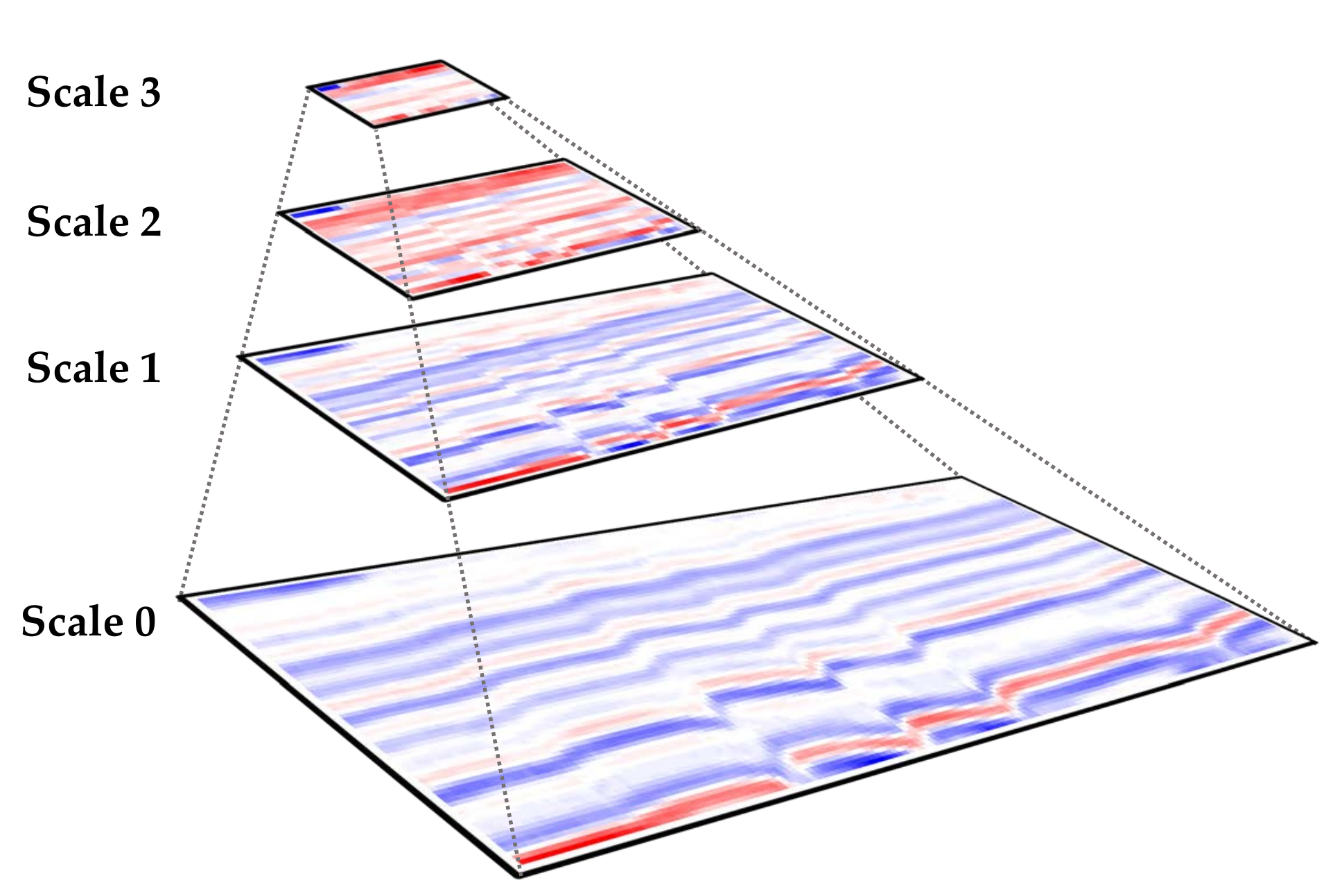}
		\caption{An illustration of a 4-scale Gaussian pyramid.}
		\label{figure:pyramid}
\end{figure}

An illustration of a 4-scale Gaussian pyramid of a seismic section is shown in Figure \ref{figure:pyramid}. Note that the dimensions of the image is reduced by a factor of two as we go up the pyramid, i.e. size of $f_i[m,n] \in \mathbb{R}^{\frac{M}{2^i}\times \frac{N}{2^i}}$. The Gaussian kernel serves as a low-pass filter and is followed by a downsampling step to omit redundant information. 

\begin{figure}[ht!]
	\centering
	\includegraphics[width =\linewidth]{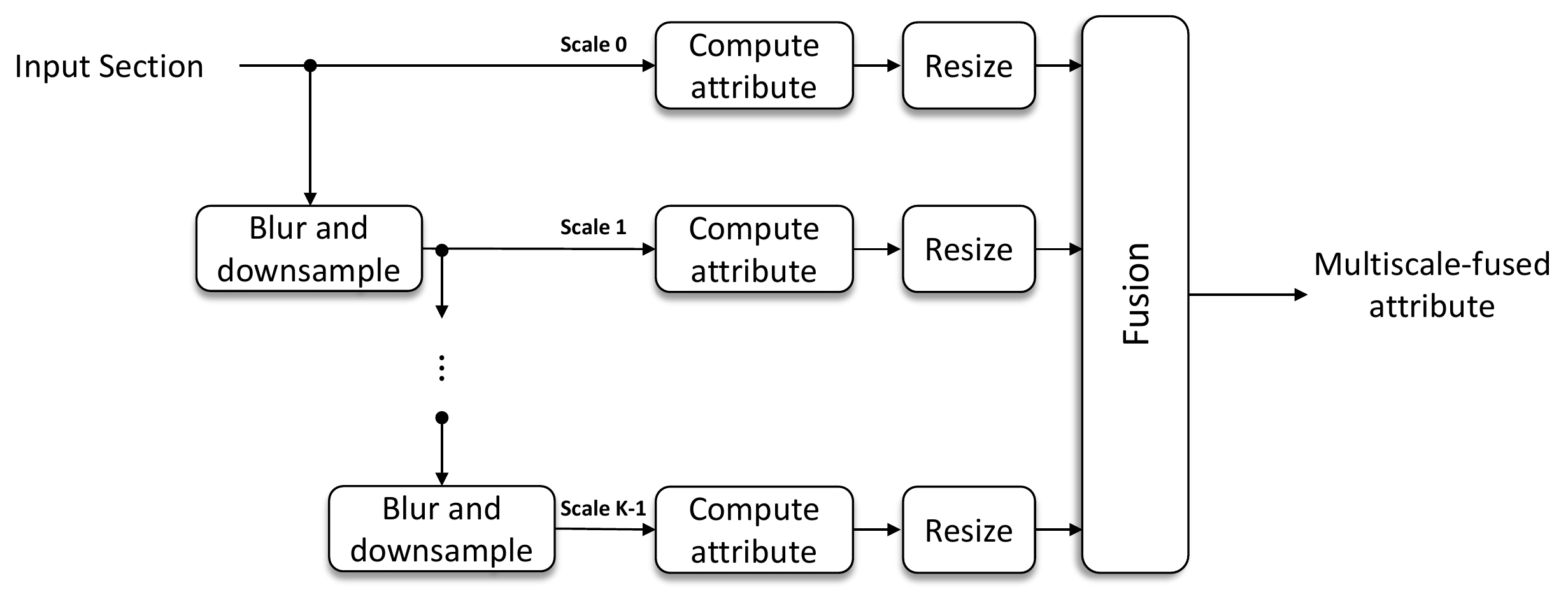}
	\caption{The proposed multiscale fusion workflow}
	\label{figure:workflow}
\end{figure}

Next, the seismic attribute is computed from each scale of the Gaussian pyramid. Thus, a $K$ scale Gaussian pyramid will produce $K$ attribute maps. Note that the attribute map at scale 0 is the attribute computed by the conventional method. It is expected that attributes at higher scales of the pyramid will have lower resolution while preserving large-scale features. On the other hand, lower scales of the pyramid produce detailed, but noisy, attributes. To produce a single enhanced attribute, a fusion mechanism is devised, which requires the attribute maps to be of the same size. Therefore, all maps are resized by means of interpolation to be of the same size as the attribute map at scale 0 prior to fusion. The proposed method is depicted in Figure \ref{figure:workflow}.

The multiscale fusion step can be interpreted as filtering in scale space as opposed to the conventional methods where filtering is done spatially. For example, if we use a $K$-level Gaussian pyramid then each sample in the input seismic section will have $K$ corresponding attribute samples, i.e. one sample from each scale. These $K$ samples are fused to obtain a single representative sample by means of simple averaging, or a weighted averaging. For instance, to emphasize features from high-resolution scales, we can use a weighted mean fusion scheme with larger weights for lower scales of the pyramid. We can also use nonlinear operators for multiscale fusion such as rank filters. An example of a rank filter that is typically used for noise reduction is the median filter. In our experiments, the median filter gives the best results because it is less sensitive to extreme values (i.e. outliers). 

\section{Results}

\begin{figure}[ht!]
	\centering
	\includegraphics[width = 0.8\linewidth]{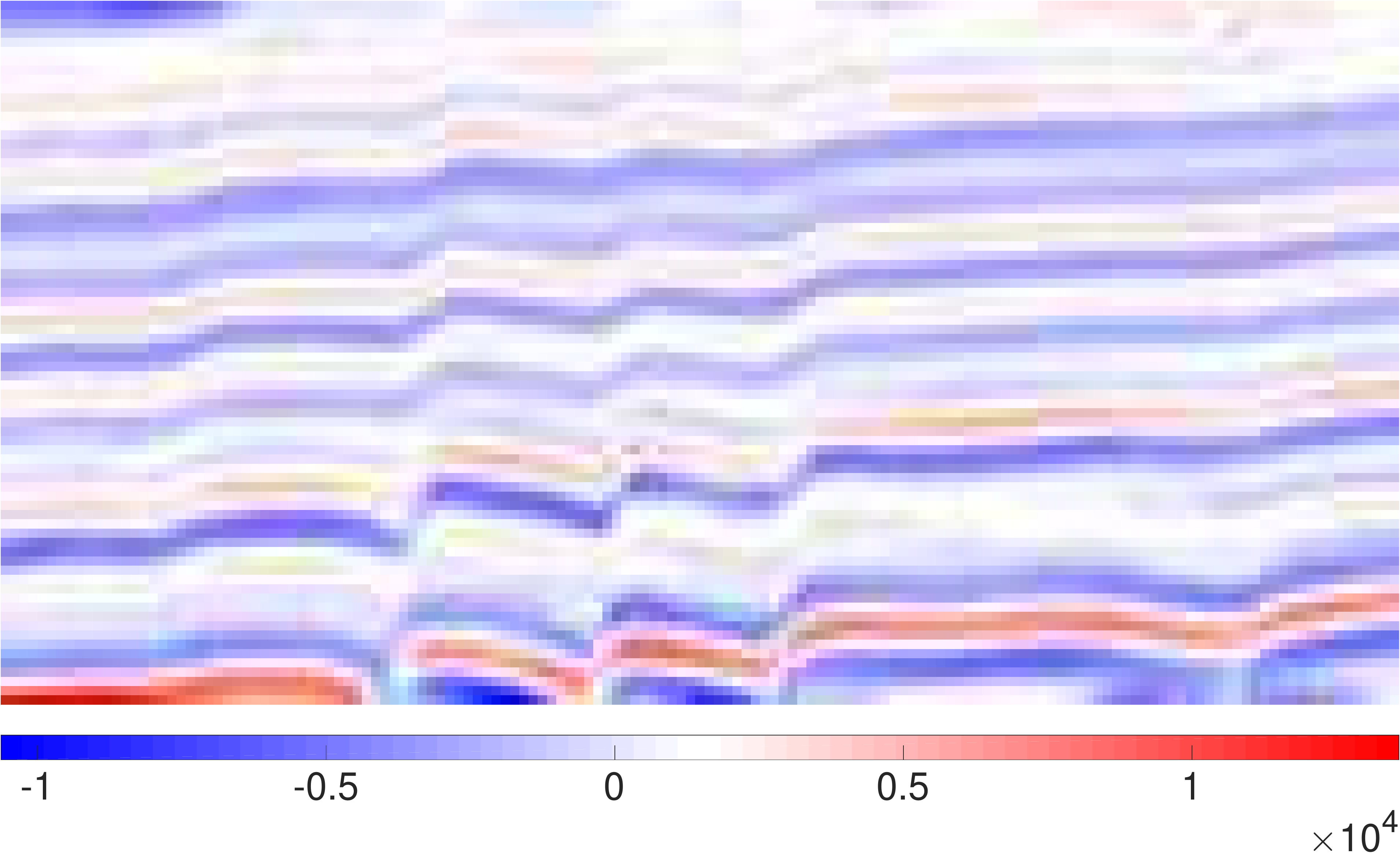}
	\caption{Seismic amplitude of inline 187 from 3D  seismic  dataset from the Great South Basin in New Zealand.}
	\label{figure:inline}
\end{figure}

\begin{figure}[ht!]
	\centering
	\subfigure[Original attribute]{\includegraphics[width = 0.8\linewidth]{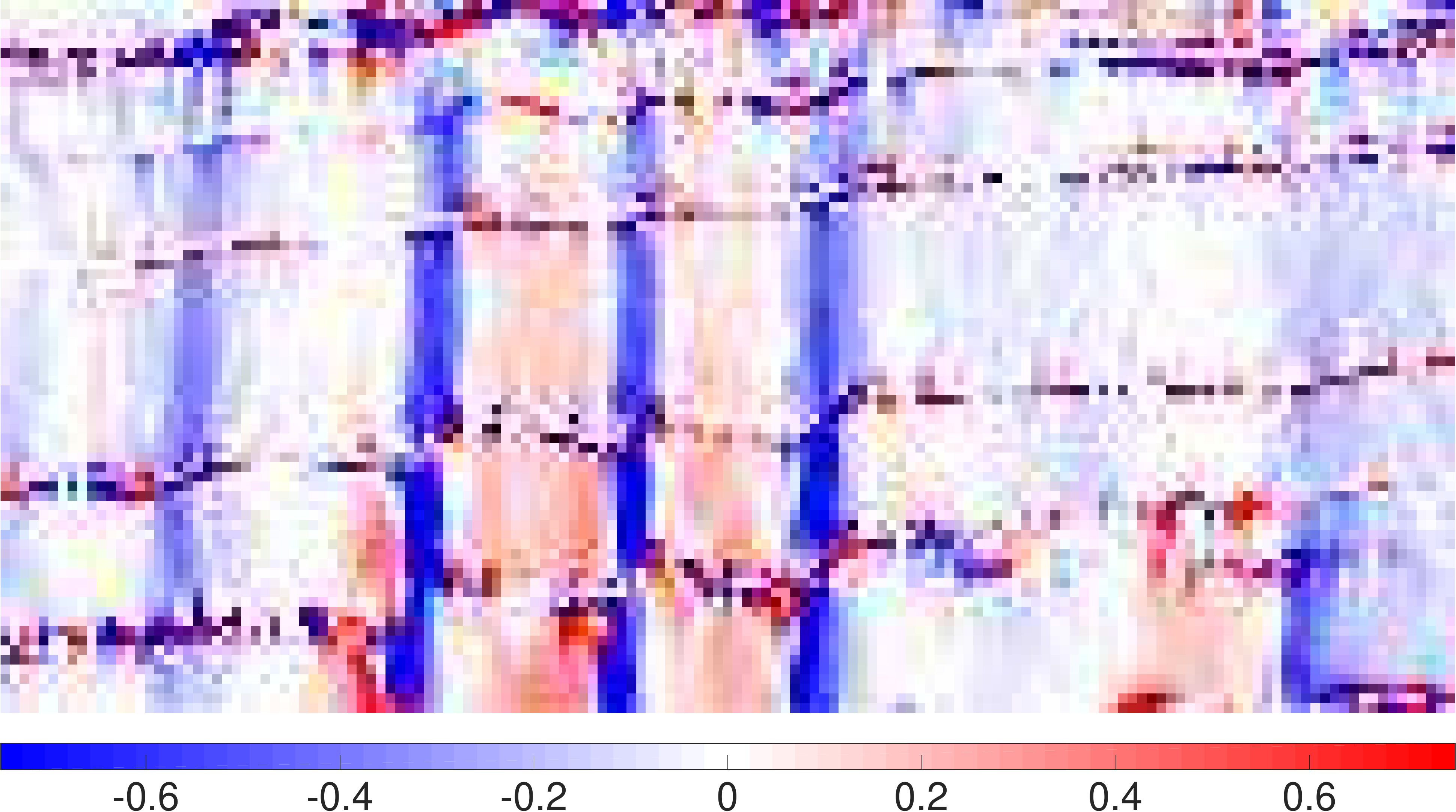}\label{figure:DipxOriginal}}
	\subfigure[Multiscale attribute with median fusion]{\includegraphics[width = 0.8\linewidth]{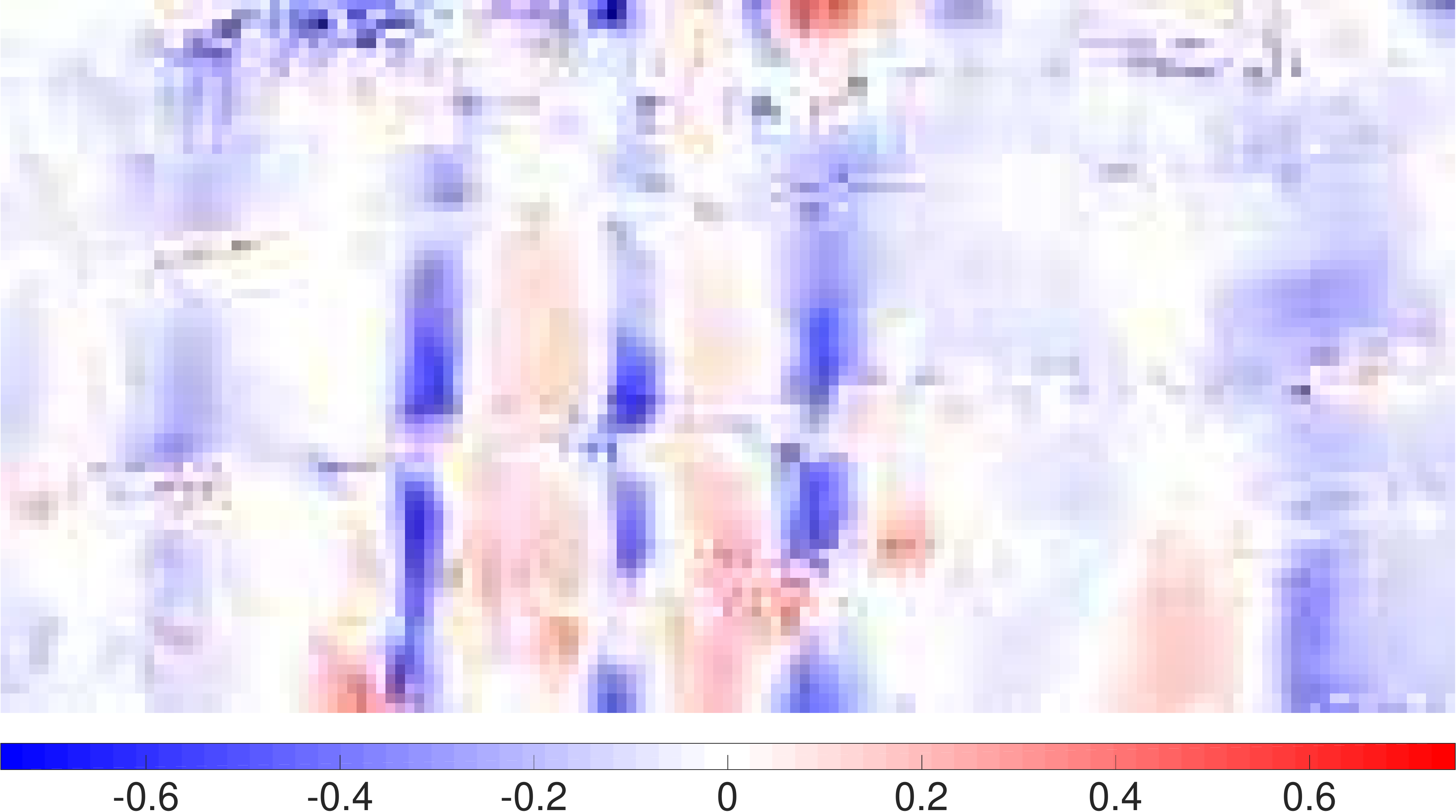}\label{figure:DipxMedian}}
	\caption{Dip attribute of seismic inline in figure \ref{figure:inline}.}
	\label{figure:DipComparison}
\end{figure}

To demonstrate its capabilities in improving the resolution of seismic geometric attributes, the proposed multiscale fusion workflow was applied to a subset of the 3D seismic dataset from the Great South Basin in New Zealand, where the dominant structural feature is polygonal faulting. We start by computing the fundamental dip attribute for inline 187 (Figure \ref{figure:inline}). The corresponding dip attribute generated by the conventional method as detailed by \cite{barnes1996theory} is shown in Figure \ref{figure:DipxOriginal}. The enhanced dip attribute computed using the proposed multiscale fusion method is shown in Figure \ref{figure:DipxMedian}. It is clear that the proposed technique helps greatly in enhancing the attribute in terms of noise reduction as well as small-scale fault detection, especially along the weak reflectors.

\begin{figure}[ht!]
	\centering
	\includegraphics[width = 0.8\linewidth]{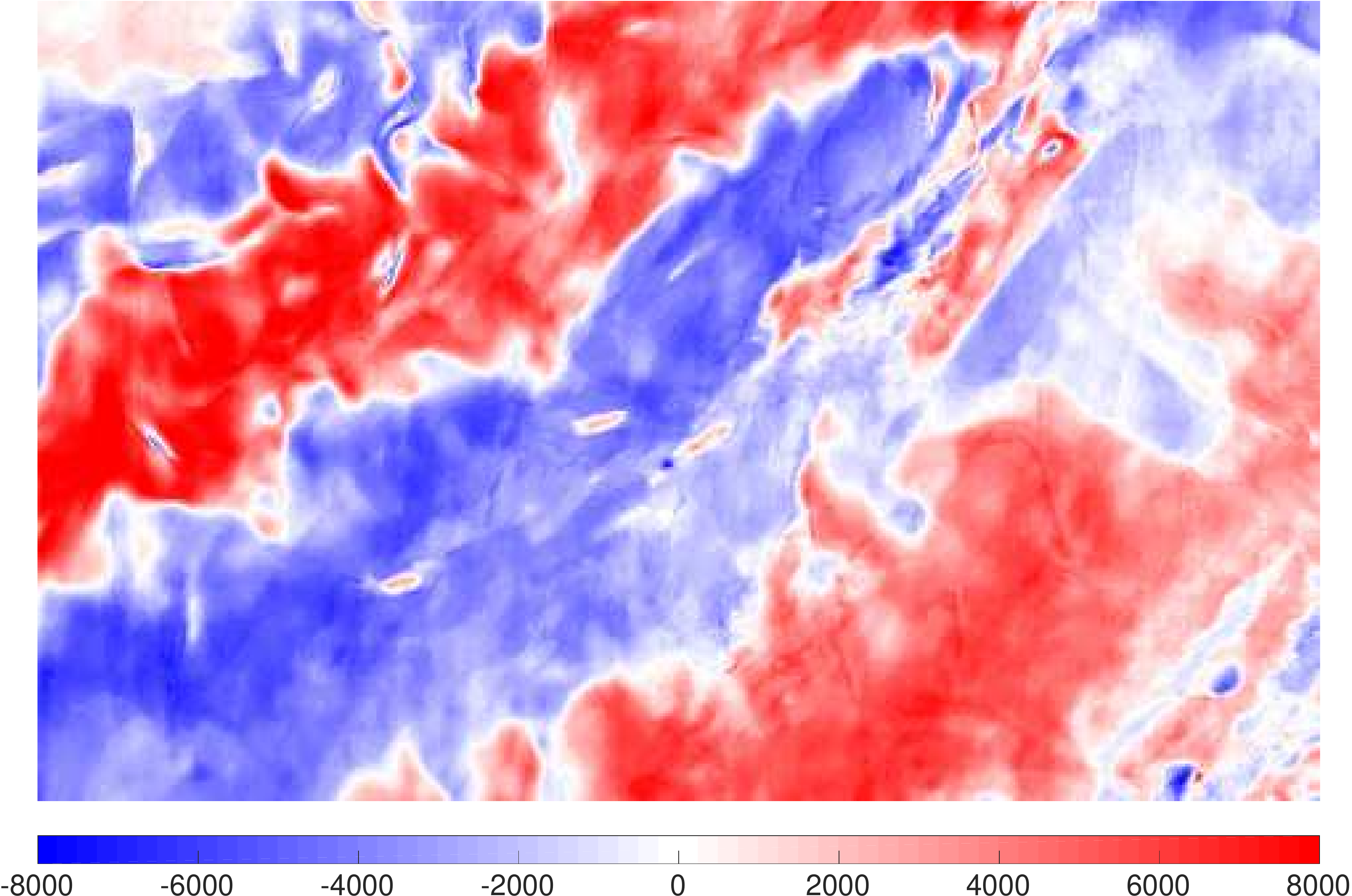}
	\caption{Seismic amplitude of the time section at 1256 ms from 3D  seismic  dataset from the Great South Basin in New Zealand.}
	\label{figure:timeslice}
\end{figure}

\begin{figure}[ht!]
	\centering
	\subfigure[Original attribute.]{\includegraphics[width = 0.8\linewidth]{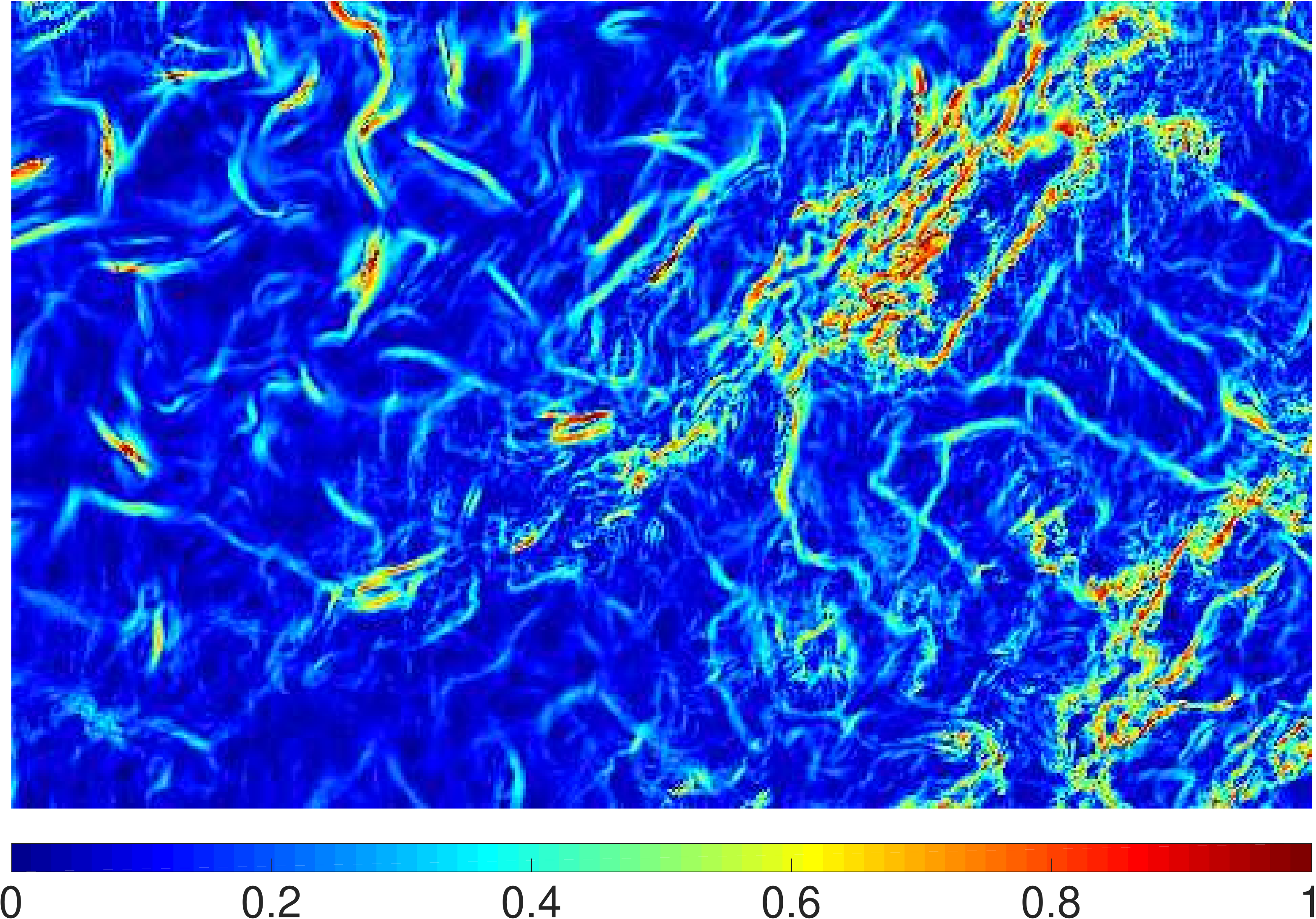}	\label{figure:DipAngle_original}}
	\subfigure[Multiscale attribute with median fusion.]{\includegraphics[width = 0.8\linewidth]{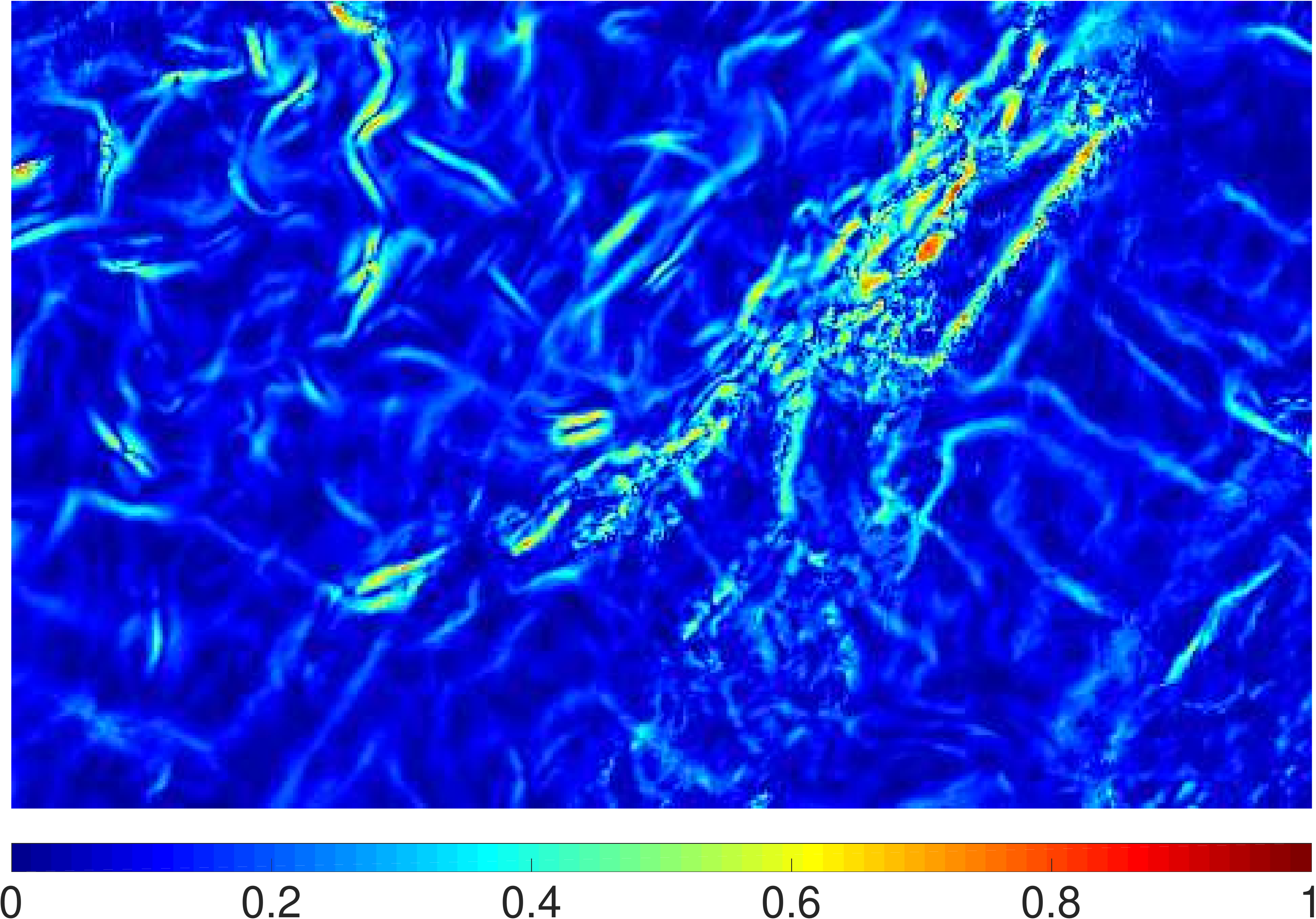}\label{figure:DipAngle_median}}
	\caption{Dip Angle attribute of the time section in Figure \ref{figure:timeslice}.}
	\label{figure:DipAngle}	
\end{figure}

Next, we illustrate the improvements of the proposed technique in highlighting the faults as observed in the time section at 1256 ms (Figure \ref{figure:timeslice}). In particular, the dip angle attribute is estimated by combining the inline dip and the crossline dip. We can see that the chaotic non-fault features are significantly removed as depicted in Figure \ref{figure:DipAngle_median}, providing a better visualization of the faults compared to the original attribute shown in Figure \ref{figure:DipAngle_original}. 

Finally, we extend the proposed method to generate high-order geometric attributes, including the most positive curvature (Figures \ref{figure:MostPositiveCurvature} and \ref{figure:MostPositiveCurvature_cropped}) and most negative curvature (Figures \ref{figure:MostNegativeCurvature} and \ref{figure:MostNegativeCurvature_cropped}), which are considered very powerful for fault detection and fracture characterization. It is evident that the proposed multiscale fusion scheme helps not only suppress the noise but also promote the identification of the curvature lineaments as faults and fractures.

\newcommand*{\Cwidth}{0.45}

\begin{figure*}[ht!]
	\centering
	\subfigure[Original attribute.]{\includegraphics[width = \Cwidth\linewidth]{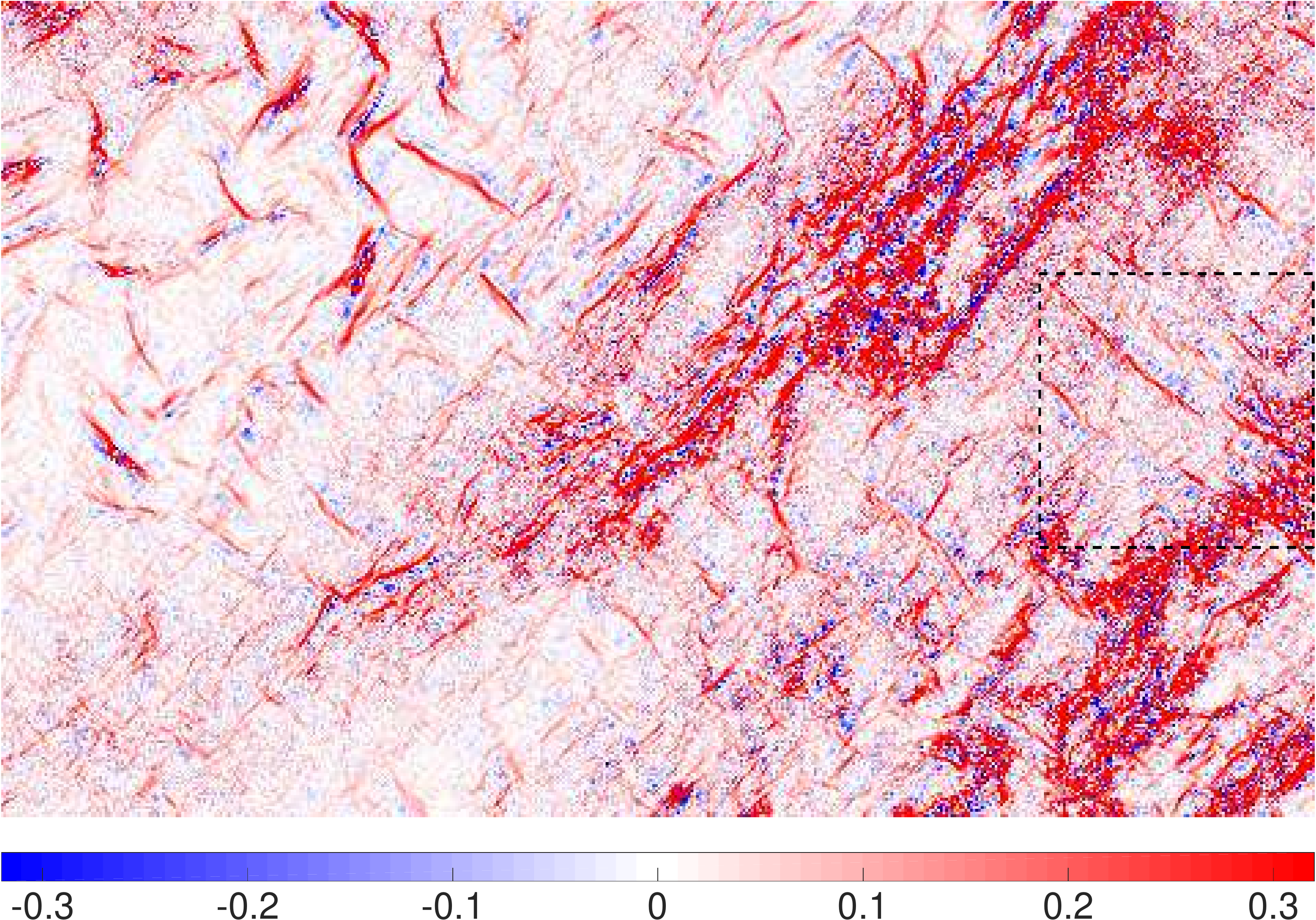}	\label{figure:Kmp_original}}
	\subfigure[Multiscale attribute with median fusion.]{\includegraphics[width = \Cwidth\linewidth]{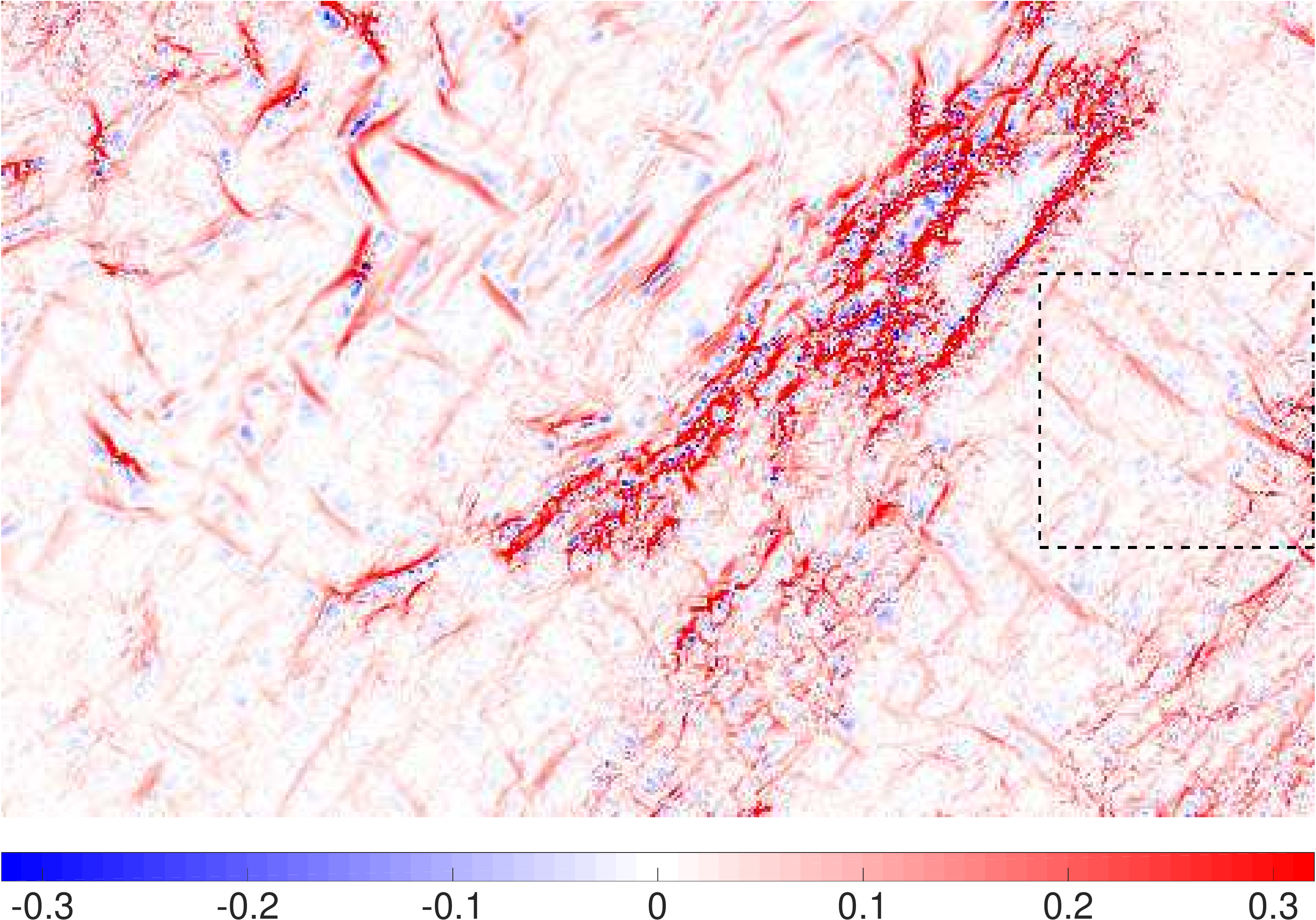}\label{figure:Kmp_median}}
	\caption{Most positive curvature attribute of the time section in Figure \ref{figure:timeslice}.}
	\label{figure:MostPositiveCurvature}	
\end{figure*}
\begin{figure*}[ht!]
	\centering
	\subfigure[Original attribute.]{\includegraphics[width = \Cwidth\linewidth]{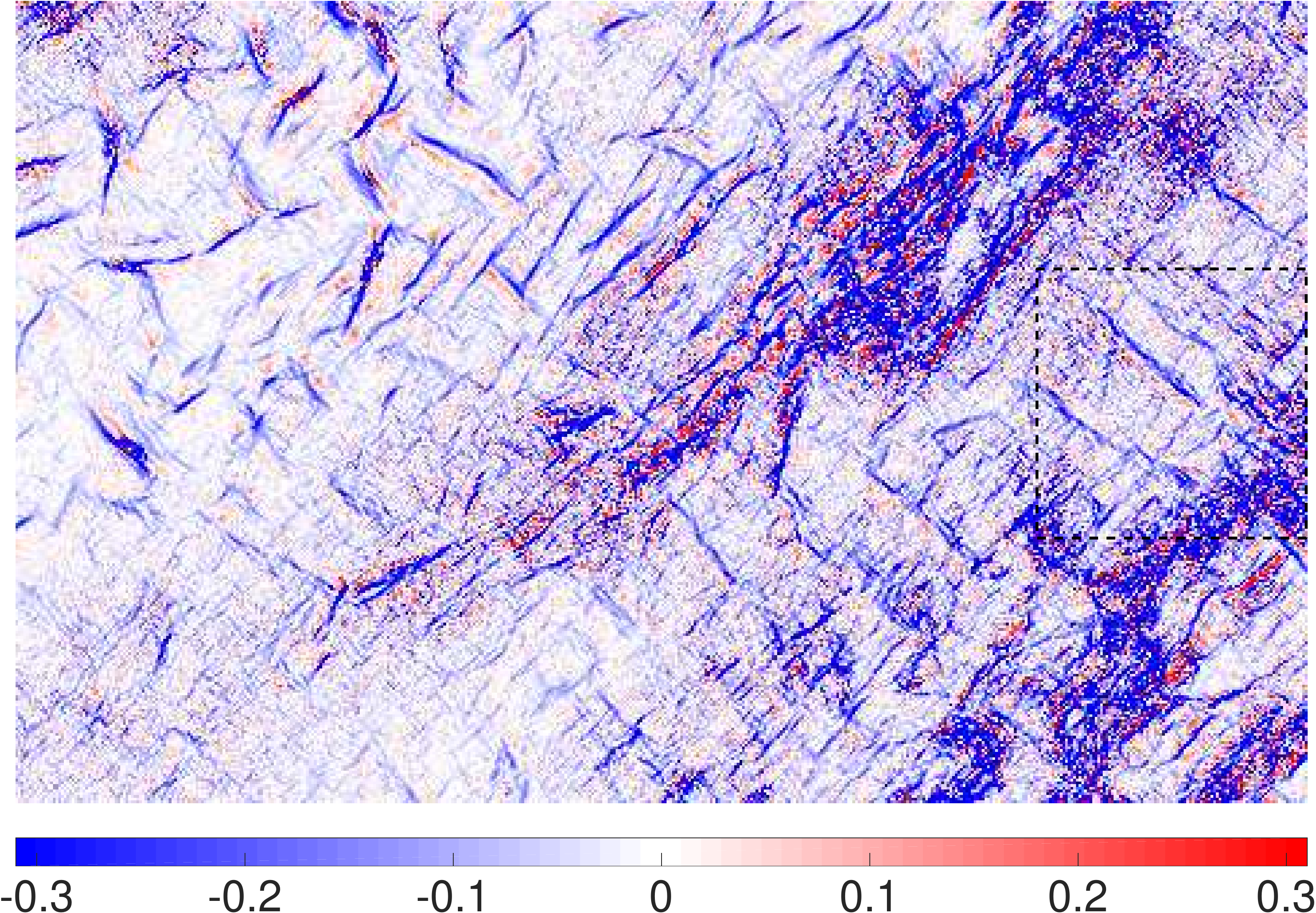}	\label{figure:Kmn_original}}
	\subfigure[Multiscale attribute with median fusion.]{\includegraphics[width = \Cwidth\linewidth]{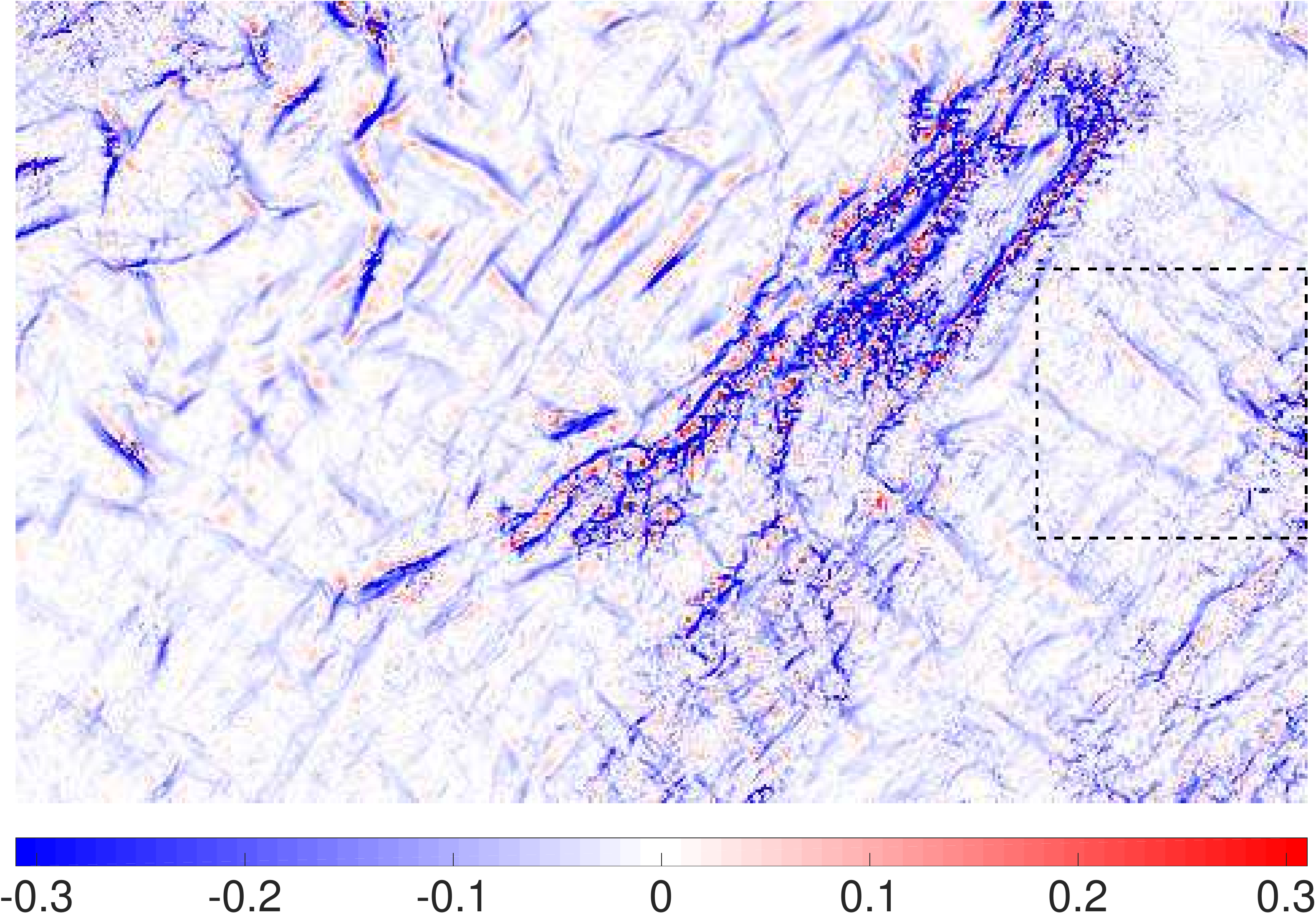}\label{figure:Kmn_median}}
	\caption{Most positive curvature attribute of the time section in Figure \ref{figure:timeslice}.}
	\label{figure:MostNegativeCurvature}	
\end{figure*}

\begin{figure}[ht!]
	\centering
	\subfigure[Original attribute.]{\includegraphics[width = 0.45\linewidth]{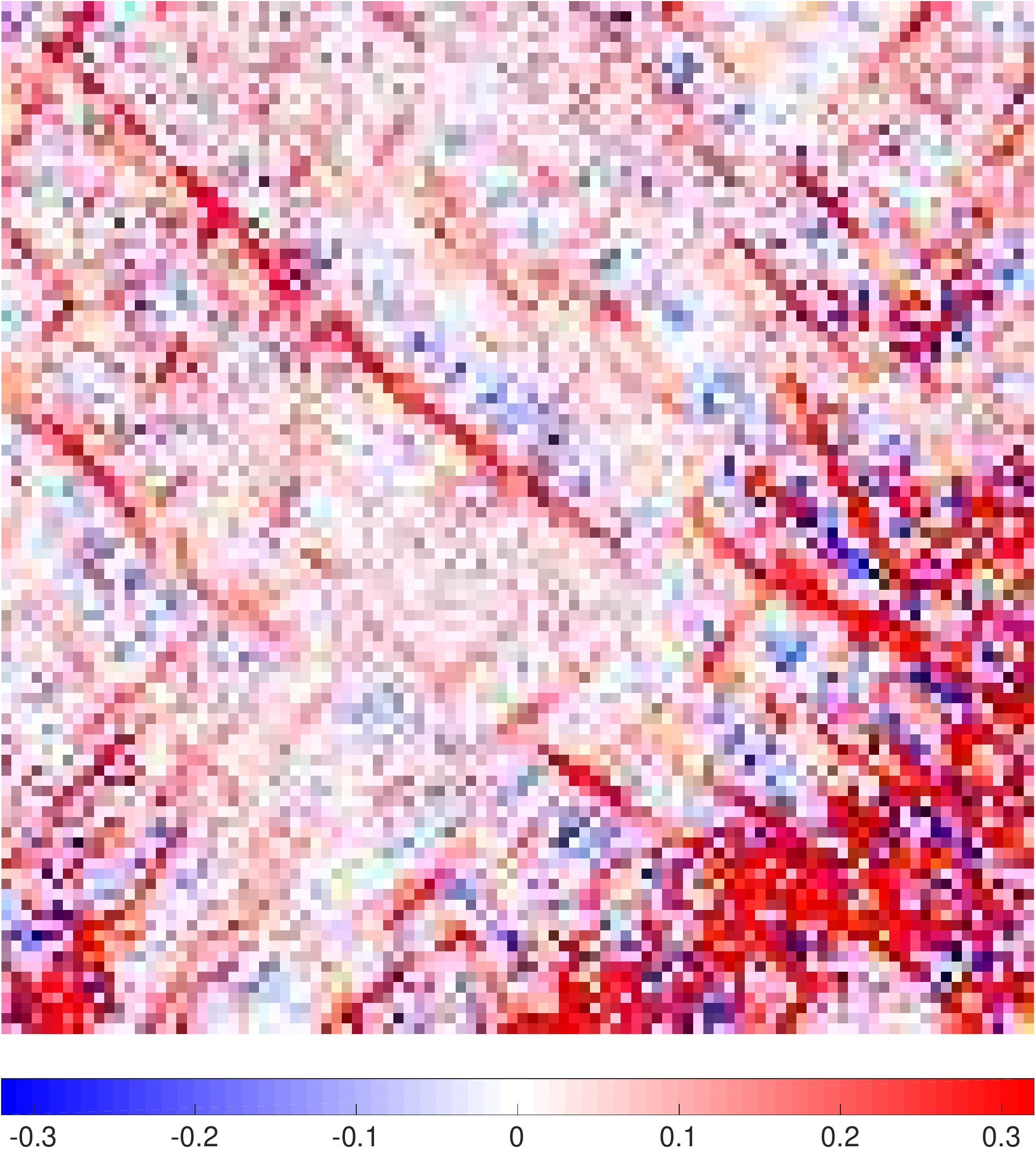}	\label{figure:Kmp_original_cropped}}
	\subfigure[Multiscale attribute with median fusion.]{\includegraphics[width = 0.45\linewidth]{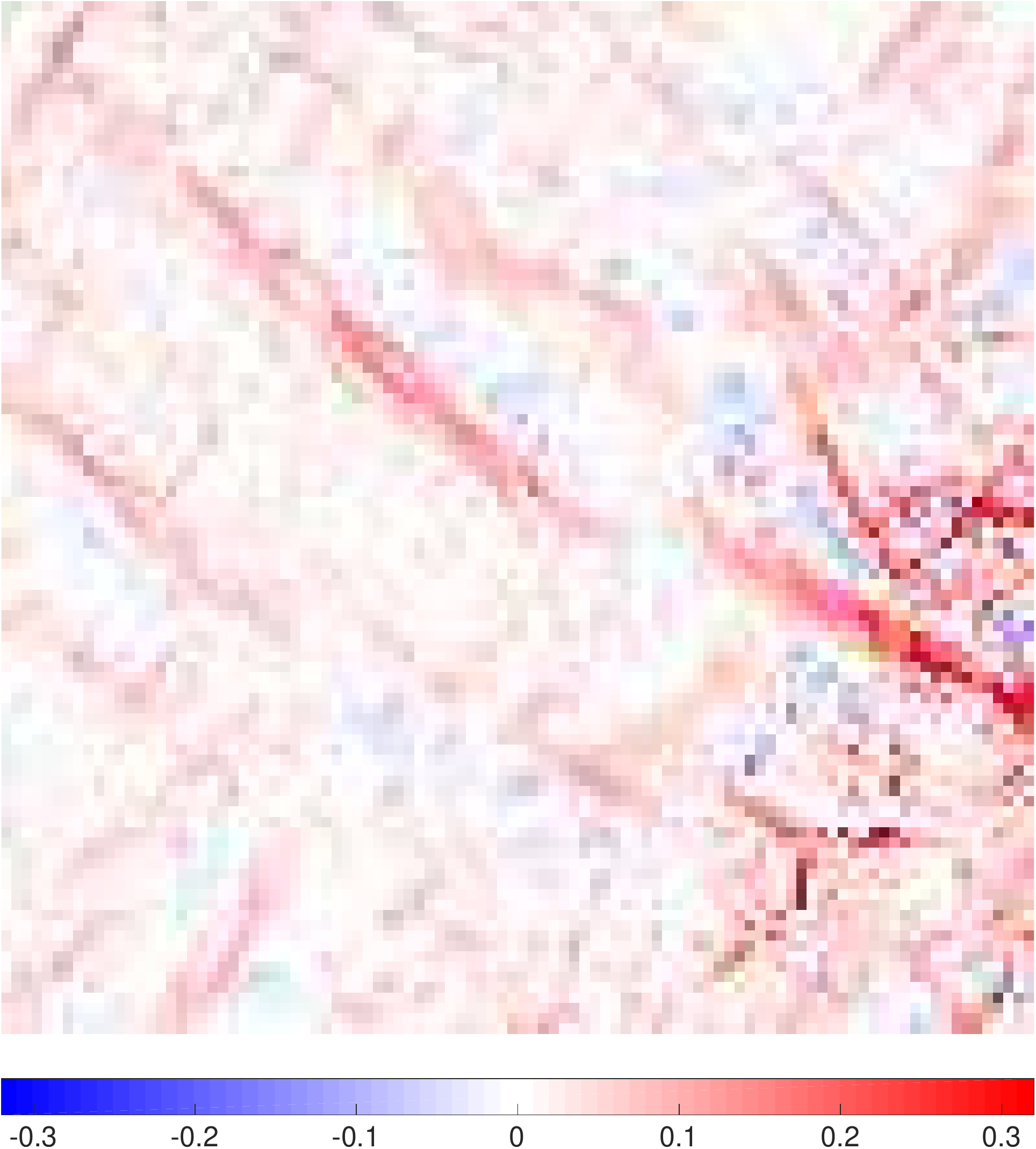}\label{figure:Kmp_median_cropped}}
    \caption{Highlighted regions in Figure \ref{figure:MostPositiveCurvature}.}
    \label{figure:MostPositiveCurvature_cropped}
    \end{figure}

\begin{figure}[ht!]
	\centering
	\subfigure[Original attribute.]{\includegraphics[width = 0.45\linewidth]{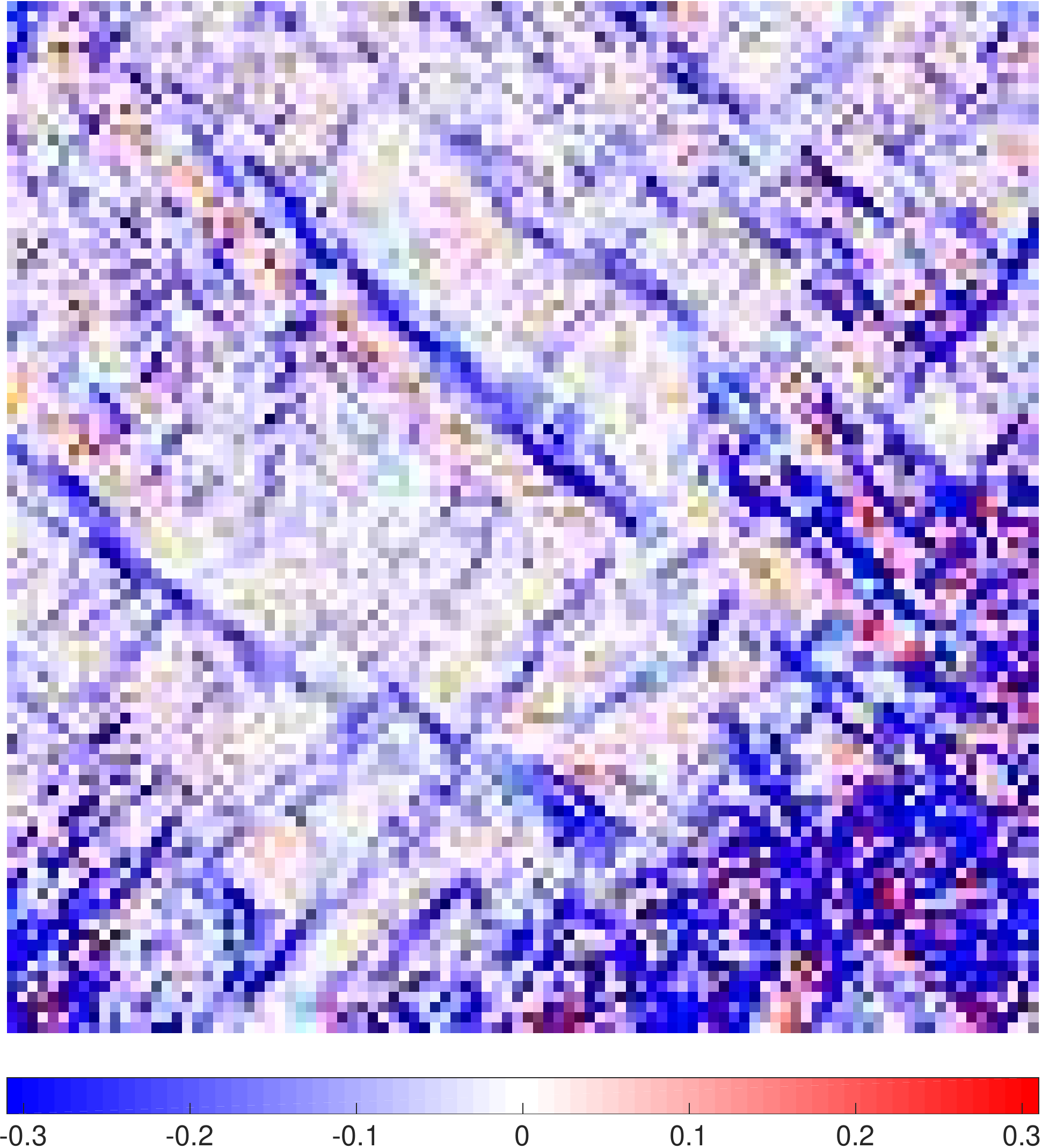}	\label{figure:Kmn_original_cropped}}
    \subfigure[Multiscale attribute with median fusion.]{\includegraphics[width = 0.45\linewidth]{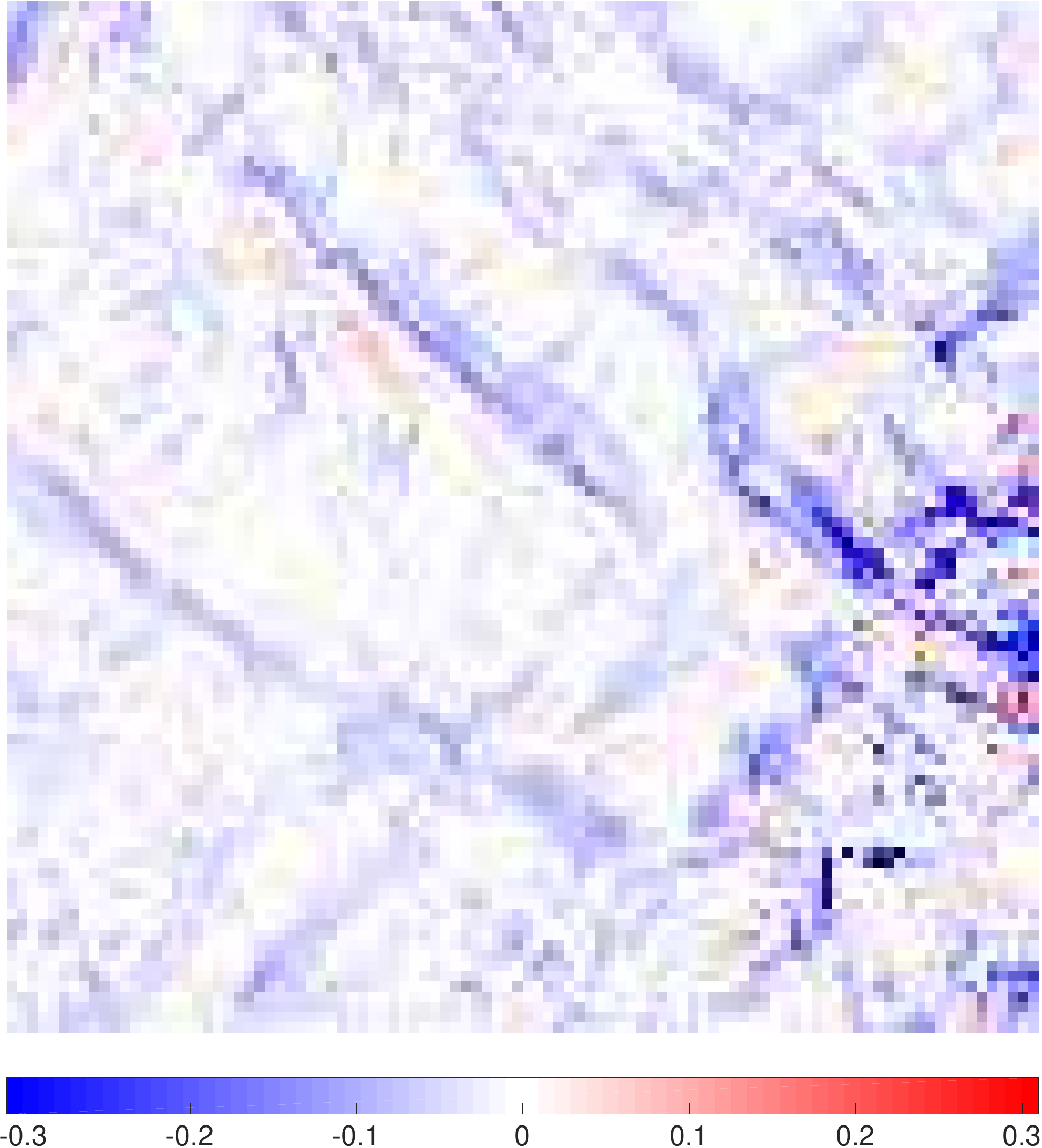}\label{figure:Kmn_median_cropped}}
    \caption{Highlighted regions in Figure \ref{figure:MostNegativeCurvature}.}
    \label{figure:MostNegativeCurvature_cropped}
\end{figure}

\section{Conclusions}
We have presented an innovative workflow, namely multiscale fusion,  for improving seismic geometric attributes, such as dip and curvature, which are known to be highly sensitive to the noise present in seismic data. Such improvement has been verified through applications to the 3D seismic dataset from New Zealand, in which the high-frequency noise was successfully suppressed while preserving small-scale seismic faults and fractures. The proposed method provides enhanced attributes that can be utilized in the identification and interpretation the polygonal faulting developed in this area. We conclude that, in addition to the dip and curvature attributes, the proposed workflow holds great potential for improving other multitrace seismic attributes, such as coherence, flexure, and GLCM.

\section{Acknowledgment}
This work is supported by the Center for Energy and Geo Processing (CeGP) at Georgia Tech and King Fahd University of Petroleum and Minerals. We thank the New Zealand Petroleum and Minerals (NZP\&M) for providing the 3D seismic dataset over the Great South Basin (GSB).

\end{document}